\def\ra{\rangle}
\def\la{\langle}
\def\be{\begin{equation}}
\def\ee{\end{equation}}
\def\ba{\begin{array}}
\def\ea{\end{array}}
\def\dps{\displaystyle}

\documentclass[aps,pre,preprint,showpacs,amsmath,amssymb,amsfonts,preprintnumbers]
{revtex4}
\usepackage{epsfig,graphicx}
\usepackage{amsmath}

\begin{document}

\baselineskip=18pt \setcounter{page}{1}
\centerline{\large\bf Local Unitary Invariants for Multipartite Quantum Systems}
\vspace{4ex}
\begin{center}

Jing Wang$^{1,2}$, Ming Li$^{2,3}$, Shao-Ming Fei$^{1,3}$ and
Xianqing Li-Jost$^{3}$

\vspace{2ex}

\begin{minipage}{6in}

{\small $~^{1}$ School of Mathematical Sciences, Capital Normal
University, 100048 Beijing, China}

{\small $~^{2}$  College of the Science, China University of
Petroleum, 266580 Qingdao, China}

{\small $~^{3}$ Max-Planck-Institute for Mathematics in the Sciences, 04103 Leipzig, Germany}

\end{minipage}
\end{center}

\begin{center}
\begin{minipage}{5in}
\vspace{1ex} \centerline{\large Abstract} \vspace{1ex}
We present an approach of constructing invariants under local unitary transformations
for multipartite quantum systems. The invariants constructed in this way can be
complement to that in [Science 340 (2013) 1205-1208]. Detailed examples are given to
compute such invariant in detail. It is shown that these invariants can be used to
detect the local unitary equivalence of degenerated quantum states.

\vfill
\smallskip
\end{minipage}\end{center}
\bigskip

As the characteristic trait of quantum theory, quantum entanglement has been
extensively studied in recent years. One approach to study the multipartite
quantum entanglement is to study the local unitary (LU) invariants of the system.
Actually, there are many characters related to the
quantum entanglement, such as the degree of entanglement
\cite{eof1,eof2}, the maximal violations of Bell inequalities
\cite{bell1,bell2,bell3,bell4} and the teleportation fidelity
\cite{tel1,tel2}. The quantities related are invariant under local unitary
transformations, while two quantum states with the
same entanglement may be not equivalent under local unitary
transformations. It is of great importance to investigate
the invariants under LU transformations.

In \cite{makhlin} the authors presented a complete set of 18
polynomial invariants for the local unitary(LU) equivalence of
two-qubit mixed states. For three qubits states, nice results have
been obtained in \cite{linden, Linden}. The authors have proposed
invariants for some generic mixed states in \cite{SFG, SFW, SFY},
tripartite pure and mixed states in \cite{SCFW}. For bipartite mixed
quantum systems, Zhou et.al \cite{zhou} have presented a complete
set of invariants such that two density matrices are locally
equivalent if and only if all these invariants have equal values in
these density matrices. In \cite{zhang1}, the authors have
investigated the LU equivalence problem in terms of matrix
realignment and partial transposition. A necessary and sufficient
criterion for the local unitary equivalence of multipartite states,
together with explicit forms of the local unitary operators have
been presented. The criterion is shown to be operational for states
having eigenvalues with multiplicity of no more than 2. Recently, we
have given a complete classification under LU operation for
multi-qubit quantum states\cite{pralu2014}, which also supplies a
numerically computable protocol to detect the LU equivalence of any
two multi-qubit mixed states. In \cite{zhang} the case for
multipartite system is studied and a complete set of invariants is
presented for a class of mixed states. The authors in \cite{pr} have
derived a set of invariants by using the partial transpose and
realignment. However, generally a complete set of LU invariants is
still missing. In this paper, we construct invariants under local
unitary transformation for multipartite quantum systems. We further
show by examples that these invariants are computable and can be
used as criterion for detecting local unitary equivalence of
degenerating quantum states.

Consider a pure quantum state $\rho=|\psi\ra\la\psi|$ in
multipartite quantum system $H_1\otimes H_2\otimes\cdots\otimes H_N$
systems with $dim\,H_i=d_i$, $i=1,2,\cdots,N$. We first recall some
results in \cite{science}. Let $\rho_i=Tr_{\hat{i}}\rho$,
$i=1,2,\cdots, N$, be the one-body reduced matrices of $\rho$, where
$Tr_{\hat{i}}$ denotes the trace over all the subsystems except the
$i$th. Denote
$S=\{\lambda_1^1,\lambda_2^1,\cdots,\lambda_{d_1}^1,\lambda_1^2,\cdots,\lambda_{d_2}^2,\cdots,
\lambda_1^N,\cdots,\lambda_{d_N}^N\}$ the set of all the eigenvalues
of $\rho_i$. Walter et. al showed that $S$ is in fact a convex
polytope which represents an entanglement class. If the collection
of eigenvalues $S$ of the one-body reduced density matrices of a
pure quantum state does not lie in an entanglement polytope, then
the state does not belong to the corresponding entanglement class.
The elements in the set $S$ are just the invariants of
$\rho=|\psi\ra\la\psi|$ under local unitary transformation. In the
following we present a set of invariants under local unitary
transformation that is complement to $S$, by using the idea in
constructing invariants of bipartite systems \cite{zhang}. The
invariants obtained in this way are independent of the detailed
spectral expressions of a density matrix.

Let $\rho_{ij}$, $1\leq i\neq j \leq N$, denote the reduced density
matrices with rank $r_{ij}$ of a pure state $\rho$, acting on
$H_i\otimes H_j$. The spectral decomposition of $\rho_{ij}$ is
represented as \be\label{spect}
\rho_{ij}=\sum_k^{r_{ij}}\Lambda_k^{ij}|X_k^{ij}\ra\la X_k^{ij}|,
\ee where $\Lambda_k^{ij}$, $k=1,2,...,r_{ij}$, are the eigenvalues
of $\rho_{ij}$, with the corresponding eigenvectors $|X_k^{ij}\ra$.
Set $|\tilde{X}_k^{ij}\ra=\sqrt{\Lambda_k^{ij}}|X_k^{ij}\ra$. One
has that $\rho_{ij}=\sum_k^{r_{ij}}|\tilde{X}_k^{ij}\ra\la
\tilde{X}_k^{ij}|$. Let $A_k^{ij}$ be the matrix with entries given
by the coefficients of the bipartite state vector
$|\tilde{X}_k^{ij}\ra$ in computational basis. Define the matrix
$\Omega^{ij}$ with entries
$(\Omega^{ij})_{lk}=Tr(A_l^{ij}(A_k^{ij})^{T})$, where $T$ stands
for the transposition of a matrix. The character polynomial of
$\Omega^{ij}$ is given by \be\label{ploy} \det\{\lambda
I-\Omega^{ij}\}=\lambda^{r_{ij}^2}+C_1^{ij}\lambda^{r_{ij}^2-1}+\cdots+C_{r_{ij}}^{ij}.
\ee

{\bf{Theorem 1:}} The coefficients $C_{\alpha}^{ij}$ in (\ref{ploy})
must be the invariants of $\rho$ under local unitary transformation,
i.e. $\{C_{\alpha}^{ij},~\alpha=1,...,r_{ij},~1\leq i \neq j \leq N\}$
form a set of invariants for $\rho=|\psi\ra\la\psi|$  under local
unitary transformations.

{\bf{Proof:}} Assume $|\psi'\ra=U_1\otimes U_2\otimes\cdots\otimes U_N|\psi\ra$. One has that
\begin{eqnarray}\label{4}
\rho_{ij}'&&=Tr_{\hat{i}\hat{j}} \rho'=Tr_{\hat{i}\hat{j}}
U_1\otimes U_2\otimes\cdots\otimes
U_N|\psi\ra\la\psi|U_1^{\dag}\otimes U_2^{\dag}\otimes\cdots\otimes
U_N^{\dag}\nonumber\\
&&=Tr_{\hat{i}\hat{j}} \rho=U_i\otimes U_j\rho_{ij}U_i^{\dag}\otimes
U_j^{\dag}.
\end{eqnarray}
Hence we have
\be
\rho_{ij}'=\sum_k^{{r'_{ij}}}|\tilde{X'}_k^{ij}\ra\la
\tilde{X'}_k^{ij}|=\sum_k^{{r_{ij}}}U_i\otimes
U_j|\tilde{X}_k^{ij}\ra\la \tilde{X}_k^{ij}|U_i^{\dag}\otimes
U_j^{\dag},
\ee
where $r'_{ij}$ is the rank of $\rho'_{ij}$.

Therefore, we have $|\tilde{X'}_k^{ij}\ra=U_i\otimes
U_j|\tilde{X}_k^{ij}\ra$. Correspondingly we derive that
${A'}_l^{ij}=U_iA_l^{ij}U_j^T$. Hence
${\Omega'}^{ij}={\Omega}^{ij}$, which leads to that the
coefficients of the characteristic polynomials of
$\Omega'^{ij}$ and $\Omega^{ij}$ are the same,
${C'}_{\alpha}^{ij}=C_{\alpha}^{ij}$. They are the invariants
under local unitary transformations. \hfill \rule{1ex}{1ex}

It is obvious that if two multipartite pure states have
different values for one or more invariants, they can not be LU equivalent.
In the following we give two examples to compute these invariants.

{\bf{Example 1:}} Consider the generalized GHZ state
$|GHZ\ra=(\cos{\theta},0,0,0,0,0,0,\sin{\theta})^T$. The two-body
reduced matrices are all of the same form:
\begin{eqnarray}
\label{trm}
\rho_{12}=\rho_{13}=\rho_{23}=\left(%
    \begin{array}{cccc}
      \cos{\theta} & 0 & 0 & 0\\
      0 & 0 & 0 & 0\\
      0 & 0 & 0 & 0\\
      0 & 0 & 0 & \sin{\theta}
     \end{array}
    \right).
\end{eqnarray}
Correspondingly,
\begin{eqnarray}
\label{trm1}
\Omega^{12}=\Omega^{13}=\Omega^{23}=\left(%
    \begin{array}{cccc}
      \sin^2{\theta} & 0 & 0 & 0\\
      0 & \cos^2{\theta} & 0 & 0\\
      0 & 0 & 0 & 0\\
      0 & 0 & 0 & 0
     \end{array}
    \right).
\end{eqnarray}
From (\ref{ploy}) we get the  LU invariants:
$C^{ij}_1=1$, $C^{ij}_2=-1$, $C^{ij}_3=\cos^2{\theta}\sin^2{\theta}$, $C^{ij}_4=0$, $ij\in\{12,13,23\}$.

{\bf{Example 2:}} The generalized W state can be written as
$|W\ra=(0,\alpha,\beta,0,\gamma,0,0,0)^T$, with
$\alpha^2+\beta^2+\gamma^2=1$. The two-body reduced matrices are of the forms:
\begin{eqnarray}
\rho_{12}=\left(%
    \begin{array}{cccc}
      \alpha^2 & 0 & 0 & 0\\
      0 & \beta^2 & \beta\gamma & 0\\
      0 & \beta\gamma & \gamma^2 & 0\\
      0 & 0 & 0 & 0\\
     \end{array}%
    \right);~~
\rho_{13}=\left(%
    \begin{array}{cccc}
      \beta^2 & 0 & 0 & 0\\
      0 & \alpha^2 & \alpha\gamma & 0\\
      0 & \alpha\gamma & \gamma^2 & 0\\
      0 & 0 & 0 & 0\\
     \end{array}%
    \right);~~
\rho_{23}=\left(%
    \begin{array}{cccc}
      \gamma^2 & 0 & 0 & 0\\
      0 & \alpha^2 & \alpha\beta & 0\\
      0 & \alpha\beta & \beta^2 & 0\\
      0 & 0 & 0 & 0\\
     \end{array}%
    \right).
\end{eqnarray}
The corresponding $\Omega^{ij}$ are given by
\begin{eqnarray}
\Omega^{12}=\left(%
    \begin{array}{cccc}
      \frac{(1-\alpha^2)^2}{\gamma^2} & 0 & 0 & 0\\
      0 & \alpha^2 & 0 & 0\\
      0 & 0 & 0 & 0\\
      0 & 0 & 0 & 0
     \end{array}%
    \right);~~
\Omega^{13}=\left(%
    \begin{array}{cccc}
     \frac{(1-\beta^2)^2}{\gamma^2} & 0 & 0 & 0\\
      0 & \beta^2 & 0 & 0\\
      0 & 0 & 0 & 0\\
      0 & 0 & 0 & 0
     \end{array}%
    \right);~~
\Omega^{23}=\left(%
    \begin{array}{cccc}
      \frac{(1-\gamma^2)^2}{\beta^2} & 0 & 0 & 0\\
      0 & \gamma^2 & 0 & 0\\
      0 & 0 & 0 & 0\\
      0 & 0 & 0 & 0
     \end{array}%
    \right).
\end{eqnarray}

Therefore we get the LU invariants of $|W\ra$:
$$
\ba{l}
\{C^{12}_{\alpha}|\alpha=1,2,3,4\}=\dps\{1,-1-\frac{\beta^2(\beta^2+\gamma^2)}{\gamma^2},\frac{\alpha^2(1-\alpha^2)}{\gamma^2}\},\\
\{C^{13}_{\alpha}|\alpha=1,2,3,4\}=\dps\{1,-1-\frac{\alpha^2(\alpha^2+\gamma^2)}{\gamma^2},\frac{\beta^2(1-\beta^2)}{\gamma^2}\},\\
\{C^{23}_{\alpha}|\alpha=1,2,3,4\}=\dps\{1,-1-\frac{\gamma^2(\beta^2+\gamma^2)}{\beta^2},\frac{\alpha^2(1-\alpha^2)}{\beta^2}\}.
\ea
$$

{\bf{Example 3:}} Consider a three-qutrit pure state:
$$
|\Psi\ra=\frac{1}{3}(1,0,0,0,1,0,0,0,1,0,x,0,0,0,x,x,0,0,0,0,x^2,x^2,0,0,0,x^2,0)^T,
$$
where $x=e^{-\frac{2i\pi}{3}}$. It is one of the maximally entangled
states, since the concurrence of $|\Psi\ra$ is the same as that of
$|GHZ\ra$. The two-body reduced matrices of $\rho=|\Psi\ra\la\Psi|$ are all the same. And the
corresponding $\Omega^{ij}$s are given by
$\Omega^{12}=\Omega^{13}=\Omega^{23}=Diag\{1/9,1/9,1/9,0,0,0,0,0,0\}.$
Therefore we obtain the invariants for all states that
are LU equivalent to $|\Psi\ra$: $C^{ij}_1=1$,
$C^{ij}_2=-1$, $C^{ij}_3=0.333333$, $C^{ij}_4=-0.037037$, and $C^{ij}_k=0$ for
$ij\in\{12,13,23\}$, $k=5,6,7,8,9$.

Generally, we can construct local unitary invariants
also from $n$-body ($2\leq n\leq N$) reduced density matrices of $\rho$.
We use the notation in \cite{siam} to define the matrix
unfolding of pure states $|\Psi\ra\in H_{i_1}\otimes
H_{i_2}\otimes\cdots\otimes H_{i_n}$ with the $k$th index as
\be
A_k\in H_{i_k}\otimes (H_{i_{k+1}}\otimes\cdots\otimes
H_{i_n}\otimes H_{i_1}\otimes\cdots\otimes H_{i_{k-1}}).
\ee
Here $A_k$
Let $A_k^{ij}$ be the matrix with entries given by the coefficients
of the bipartite state vector $|\tilde{X}_k^{ij}\ra$ in computational basis.
here is a $d_{i_k}\times (d_{i_{k+1}}\times\cdots\times
d_{i_n}\times d_{i_1}\times\cdots\times d_{i_{k-1}})$ matrix.

For $\rho=|\psi\ra\la\psi|\in
H_1\otimes H_2\otimes\cdots\otimes H_N$, the $k$-body
reduced matrices $\rho_{i_1i_2\cdots i_k}$ are given by by tracing over all the
subsystems except $H_{i_1}H_{i_2}\cdots H_{i_k}$,
$\rho_{i_1i_2\cdots i_k}=Tr_{\hat{i_1}\cdots \hat{i_k}}\rho$. Denote
$r_I$ the rank of $\rho_{i_1i_2\cdots i_k}$ and
$I=\{i_1i_2\cdots i_k\}$. Let
$\rho_I=\sum_m^{r_I}\Lambda_m^{I}|X_m^I\ra\la X_m^I|$ be the
spectral decomposition of $\rho_I$. Set $|\tilde{X}_m^I\ra=\sqrt{\Lambda_m^I}|X_m^I\ra$.
Thus
\be
\rho_I=\sum_m^r|\tilde{X}_m^I\ra\la \tilde{X}_m^I|.
\ee
By using the matrix unfolding of multi-tensor one can represent
$|\tilde{X}_m^I\ra$ in the matrix form $(A_x^I)_{m}$ with $x\in
I$. There are totally $k$ matrix unfolding forms of
$|\tilde{X}_m^I\ra$. Let $\Omega_x^I$
denote the matrix with entries given by $(\Omega_x^I)_{mn}=Tr((A_x^I)_{m}(A_x^I)_{n}^T)$.

{\bf{Theorem 2:}} The coefficients $(C_{\alpha}^I)_x$ of the
character polynomial of $\Omega_x^I$,
\be\label{ploy1}
\det\{\lambda
I-\Omega_x^I\}=\lambda^{r_I^2}+(C_x^I)_1\lambda^{r_I^2-1}+\cdots+(C_x^I)_{r_I^2},
\ee
$1\leq x\leq r_I^2$, $1\leq \alpha\leq k$, are the
invariants of $\rho=|\psi\ra\la\psi|$ under local unitary transformations.

{\bf{Proof:}} Let $|\psi'\ra=U_1\otimes U_2\otimes\cdots\otimes
U_N|\psi\ra$. Similar to (\ref{4}) one has that
\begin{eqnarray}\rho_{I}'&&=Tr_{\hat{I}} \rho'=Tr_{\hat{I}}
U_1\otimes U_2\otimes\cdots\otimes
U_N|\psi\ra\la\psi|U_1^{\dag}\otimes U_2^{\dag}\otimes\cdots\otimes
U_N^{\dag}\nonumber\\
&&=Tr_{\hat{I}} \rho=U_{i_1}\otimes U_{i_2}\otimes\cdots\otimes
U_{i_k}\rho_{I}U_{i_1}^{\dag}\otimes
U_{i_2}^{\dag}\otimes\cdots\otimes U_{i_k}^{\dag}.\end{eqnarray}

From the spectral decomposition
$\rho_{I}=\sum_x|\tilde{X}_x^{I}\ra\la \tilde{X}_x^{I}|$ we derive that
\begin{eqnarray}
\rho_{I}'=\sum_x|\tilde{X'}_x^{I}\ra\la
\tilde{X'}_x^{I}|=\sum_xU_{i_1}\otimes
U_{i_2}\otimes\cdots\otimes U_{i_k}|\tilde{X}_x^{I}\ra\la
\tilde{X'}_x^{I}|U_{i_1}^{\dag}\otimes
U_{i_2}^{\dag}\otimes\cdots\otimes U_{i_k}^{\dag}.
\end{eqnarray}
Namely, $|\tilde{X'}_x^{I}\ra=U_{i_1}\otimes
U_{i_2}\otimes\cdots\otimes U_{i_k}|\tilde{X}_x^{I}\ra$. As
$(\Omega_{\alpha})^{I}_{lm}=Tr[(A_{\alpha})_l^{I}((A_{\alpha})_m^{I})^{T}]$
and
${(\Omega'_{\alpha})}^{I}_{lm}=Tr({A'_{\alpha}}_l^{I}({A'_{\alpha}}_m^{I})^{T})$
with $(A_{\alpha})_l^{I}$ and ${A'_{\alpha}}_l^{I}$ the
$\alpha$th matrix representation of $|\tilde{X}_l^{I}\ra$ and
$|\tilde{X'}_l^{I}\ra$, respectively. One gets that
${A'_{\alpha}}_l^{I}=U_{i_\alpha}{A_{\alpha}}_l^{I}U_{\hat{i_\alpha}}^T$ with
$U_{\hat{i_\alpha}}=U_{i_1}\otimes\cdots\otimes
U_{i_{\alpha}-1}\otimes U_{i_{\alpha}+1}\otimes\cdots\otimes U_{i_k}$. Then we get
${\Omega'}_{\alpha}^{I}=\Omega_{\alpha}^{I}$ which illustrates the
equivalence of the coefficients of characteristic polynomials, i.e.
$((C')_{\alpha}^{I})_x=(C_{\alpha}^{I})_x$. \hfill \rule{1ex}{1ex}

When two multipartite density matrices have degenerated eigenvalues,
it becomes a challenging problem to judge their LU equivalence.
The invariants derived in our Theorem 1 and 2 can be
used to detect the LU equivalent problem for multipartite
degenerated quantum states.

{\bf{Example 4:}} Let us consider two three-qutrit mixed quantum states:
$$
\rho=\frac{1}{2}|\psi_+\ra\la\psi_+|+\frac{1}{54}\sum_{i,j=0}^2|0ij\ra\la
0ij|+\frac{1}{81}\sum_{i,j=0}^2|1ij\ra\la
1ij|+\frac{2}{81}\sum_{i,j=0}^2|2ij\ra\la 2ij|;
$$
and
$$\sigma=\frac{1}{2}|\phi_+\ra\la\phi_+|+\frac{1}{54}\sum_{i,j=0}^2|0ij\ra\la
0ij|+\frac{1}{81}\sum_{i,j=0}^2|1ij\ra\la
1ij|+\frac{2}{81}\sum_{i,j=0}^2|2ij\ra\la 2ij|;
$$
where
$|\psi_+\ra=\frac{1}{\sqrt{3}}(|000\ra+|111\ra+|222\ra)$ and
$|\phi_+\ra=\frac{1}{\sqrt{3}}(|001\ra+|111\ra+|222\ra)$.

$\sigma$ and $\rho$ have the same eigenvalues: three non-degenerated ones $0.51857$,
$0.02206$, $ 0.01493$, and three eigenvalues $0.02469$, $0.01852$, $0.01235$ with
multiplicity 8 each. The criteria in \cite{bliu,jpa} becomes less
operational for such degenerated states.
From our Theorem we have that the coefficients $C_{\alpha}^{12}$
corresponding to $\rho_{12}$ are
$\{0, 0, 0, 0.00006, -0.00107, 0.01269, -0.09385, 0.41564, -1, 1\}$,
which is different from that from $\sigma_{12}$: $\{0, 0, 0, 0.00001, -0.00185, 0.02246, -0.16676,
0.7079, -1.46571, 1\}$.
Therefore  by using our LU invariants
we can easily conclude that $\rho$ and $\sigma$ are not equivalent under LU
transformations.

To classify quantum states under local unitary transformations
is a fundamental problem in the theory of quantum entanglement and
correlations. We have introduced sets of invariants under LU transformations derived
from the reduced matrices.
Invariants from hyperdeterminants have been also constructed in \cite{zhang}. However,
hyperdeterminants become quite difficult to compute for systems except for three-qubit ones.
The invariants proposed in this manuscript is easy to compute.
Moreover, our invariants can be used to detect the LU
equivalence of multipartite degenerated mixed states.
Since for two multipartite mixed states $\rho$ and $\rho'$,
if they are are LU equivalent, the corresponding
reduced density matrices must be also LU equivalent. Therefore
our LU invariants give rise to necessary conditions of LU equivalence for
multipartite mixed states too.

\bigskip
\noindent{\bf Acknowledgments}\, \, This work is supported by the
NSFC 11105226, 11275131; the Fundamental Research Funds for the
Central Universities No. 12CX04079A, No. 24720122013; Research Award
Fund for outstanding young scientists of Shandong Province
No. BS2012DX045.

\end{document}